\documentclass[aps,tightenlines,superscriptaddress,twocolumn]{revtex4-1}
\usepackage{graphicx}
\usepackage{bm}
\usepackage{times}
\usepackage{lineno}
\usepackage[pscoord]{eso-pic}

\newcommand{\ems}{\sqrt{s_{\rm NN}}}
\def\Journal#1#2#3#4{{#1} {\bf #2}, #3 (#4)}

\def\NIMA{Nucl. Instrum. Methods A}
\def\NPA{Nucl. Phys. A}
\def\PRL{Phys. Rev. Lett.}
\def\PRC{Phys. Rev. C}

\def\PLB{Phys. Lett. B}
\def\EPJC{Eur. Phys. J. C}

\begin{document}

\title{Light Nuclei Collectivity from $\sqrt{s_{\rm NN}}$ = 3 GeV Au+Au
Collisions at RHIC}
\date{\today}

\affiliation{Abilene Christian University, Abilene, Texas   79699}
\affiliation{AGH University of Science and Technology, FPACS, Cracow 30-059, Poland}
\affiliation{Alikhanov Institute for Theoretical and Experimental Physics NRC "Kurchatov Institute", Moscow 117218, Russia}
\affiliation{Argonne National Laboratory, Argonne, Illinois 60439}
\affiliation{American University of Cairo, New Cairo 11835, New Cairo, Egypt}
\affiliation{Brookhaven National Laboratory, Upton, New York 11973}
\affiliation{University of Calabria \& INFN-Cosenza, Italy}
\affiliation{University of California, Berkeley, California 94720}
\affiliation{University of California, Davis, California 95616}
\affiliation{University of California, Los Angeles, California 90095}
\affiliation{University of California, Riverside, California 92521}
\affiliation{Central China Normal University, Wuhan, Hubei 430079 }
\affiliation{University of Illinois at Chicago, Chicago, Illinois 60607}
\affiliation{Creighton University, Omaha, Nebraska 68178}
\affiliation{Czech Technical University in Prague, FNSPE, Prague 115 19, Czech Republic}
\affiliation{Technische Universit\"at Darmstadt, Darmstadt 64289, Germany}
\affiliation{ELTE E\"otv\"os Lor\'and University, Budapest, Hungary H-1117}
\affiliation{Frankfurt Institute for Advanced Studies FIAS, Frankfurt 60438, Germany}
\affiliation{Fudan University, Shanghai, 200433 }
\affiliation{University of Heidelberg, Heidelberg 69120, Germany }
\affiliation{University of Houston, Houston, Texas 77204}
\affiliation{Huzhou University, Huzhou, Zhejiang  313000}
\affiliation{Indian Institute of Science Education and Research (IISER), Berhampur 760010 , India}
\affiliation{Indian Institute of Science Education and Research (IISER) Tirupati, Tirupati 517507, India}
\affiliation{Indian Institute Technology, Patna, Bihar 801106, India}
\affiliation{Indiana University, Bloomington, Indiana 47408}
\affiliation{Institute of Modern Physics, Chinese Academy of Sciences, Lanzhou, Gansu 730000 }
\affiliation{University of Jammu, Jammu 180001, India}
\affiliation{Joint Institute for Nuclear Research, Dubna 141 980, Russia}
\affiliation{Kent State University, Kent, Ohio 44242}
\affiliation{University of Kentucky, Lexington, Kentucky 40506-0055}
\affiliation{Lawrence Berkeley National Laboratory, Berkeley, California 94720}
\affiliation{Lehigh University, Bethlehem, Pennsylvania 18015}
\affiliation{Max-Planck-Institut f\"ur Physik, Munich 80805, Germany}
\affiliation{Michigan State University, East Lansing, Michigan 48824}
\affiliation{National Research Nuclear University MEPhI, Moscow 115409, Russia}
\affiliation{National Institute of Science Education and Research, HBNI, Jatni 752050, India}
\affiliation{National Cheng Kung University, Tainan 70101 }
\affiliation{Nuclear Physics Institute of the CAS, Rez 250 68, Czech Republic}
\affiliation{Ohio State University, Columbus, Ohio 43210}
\affiliation{Institute of Nuclear Physics PAN, Cracow 31-342, Poland}
\affiliation{Panjab University, Chandigarh 160014, India}
\affiliation{Pennsylvania State University, University Park, Pennsylvania 16802}
\affiliation{NRC "Kurchatov Institute", Institute of High Energy Physics, Protvino 142281, Russia}
\affiliation{Purdue University, West Lafayette, Indiana 47907}
\affiliation{Rice University, Houston, Texas 77251}
\affiliation{Rutgers University, Piscataway, New Jersey 08854}
\affiliation{Universidade de S\~ao Paulo, S\~ao Paulo, Brazil 05314-970}
\affiliation{University of Science and Technology of China, Hefei, Anhui 230026}
\affiliation{Shandong University, Qingdao, Shandong 266237}
\affiliation{Shanghai Institute of Applied Physics, Chinese Academy of Sciences, Shanghai 201800}
\affiliation{Southern Connecticut State University, New Haven, Connecticut 06515}
\affiliation{State University of New York, Stony Brook, New York 11794}
\affiliation{Instituto de Alta Investigaci\'on, Universidad de Tarapac\'a, Arica 1000000, Chile}
\affiliation{Temple University, Philadelphia, Pennsylvania 19122}
\affiliation{Texas A\&M University, College Station, Texas 77843}
\affiliation{University of Texas, Austin, Texas 78712}
\affiliation{Tsinghua University, Beijing 100084}
\affiliation{University of Tsukuba, Tsukuba, Ibaraki 305-8571, Japan}
\affiliation{United States Naval Academy, Annapolis, Maryland 21402}
\affiliation{Valparaiso University, Valparaiso, Indiana 46383}
\affiliation{Variable Energy Cyclotron Centre, Kolkata 700064, India}
\affiliation{Warsaw University of Technology, Warsaw 00-661, Poland}
\affiliation{Wayne State University, Detroit, Michigan 48201}
\affiliation{Yale University, New Haven, Connecticut 06520}

\author{M.~S.~Abdallah}\affiliation{American University of Cairo, New Cairo 11835, New Cairo, Egypt}
\author{B.~E.~Aboona}\affiliation{Texas A\&M University, College Station, Texas 77843}
\author{J.~Adam}\affiliation{Brookhaven National Laboratory, Upton, New York 11973}
\author{L.~Adamczyk}\affiliation{AGH University of Science and Technology, FPACS, Cracow 30-059, Poland}
\author{J.~R.~Adams}\affiliation{Ohio State University, Columbus, Ohio 43210}
\author{J.~K.~Adkins}\affiliation{University of Kentucky, Lexington, Kentucky 40506-0055}
\author{G.~Agakishiev}\affiliation{Joint Institute for Nuclear Research, Dubna 141 980, Russia}
\author{I.~Aggarwal}\affiliation{Panjab University, Chandigarh 160014, India}
\author{M.~M.~Aggarwal}\affiliation{Panjab University, Chandigarh 160014, India}
\author{Z.~Ahammed}\affiliation{Variable Energy Cyclotron Centre, Kolkata 700064, India}
% ...

\collaboration{STAR Collaboration}\noaffiliation


\begin{abstract}
In high-energy heavy-ion collisions, partonic collectivity is evidenced by the
constituent
quark number scaling of elliptic flow anisotropy for identified hadrons. 
A breaking of this scaling and dominance of baryonic interactions is found 
for identified hadron collective flow measurements in $\sqrt{s_{\rm NN}}$ = 3 GeV Au+Au collisions.
In this paper, we report
measurements of the first- and second-order azimuthal anisotropic
parameters, $v_1$ and $v_2$, of light nuclei ($d$, $t$, $^{3}$He, $^{4}$He)
produced in $\sqrt{s_{\rm NN}}$ = 3 GeV Au+Au collisions at the STAR experiment.  An
atomic mass number scaling is found in the measured $v_1$ slopes of light
nuclei at mid-rapidity. For the
measured $v_2$ magnitude, a strong rapidity dependence is observed. Unlike $v_2$
at higher collision energies, the $v_2$ values at mid-rapidity for all light nuclei are
negative and no scaling is observed with the atomic mass number. Calculations
by the Jet AA Microscopic Transport Model (JAM), with baryonic mean-field plus
nucleon coalescence, are in
good agreement with our observations, implying baryonic interactions
dominate the collective dynamics in 3 GeV Au+Au collisions at RHIC.  
\end{abstract}

\maketitle

\hyphenpenalty=700
\tolerance=100
\section{Introduction}
Collective motion of particle emission in high-energy heavy-ion collisions,
often referred to as collective flow, is a general
phenomenon observed over a wide range of collision energies.
The flow anisotropy parameters, $v_n$ (where $n$ represents the $n$-th harmonic order), are used to describe the azimuthal anisotropies 
in particle momentum distributions with respect to the reaction plane~\cite{prc58_1671}. 
The first- and second-order azimuthal anisotropies, $v_1$ and $v_2$, are important
probes of nuclear matter. In high energy collisions at the top RHIC
and LHC energies, they provide information on the collective hydrodynamic
expansion and transport properties of the produced Quark Gluon Plasma (QGP), while at lower collision energies of the order of a few GeV, they are sensitive to the  compressibility of the nuclear matter and nuclear equation of state~\cite{annrevnphy_63_123,science.1078070}.
The collision-energy dependence of $v_1$ and $v_2$ for different particle
species has been observed
experimentally~\cite{prl112_162301,prl105_252302}, 
and provides valuable information on the dynamical evolution of 
the strongly interacting matter. 

At high LHC energies, significant $v_2$ and $v_3$ values are reported for $d$~\cite{prc102_055203,epjc77_658}. 
In parallel and at lower energies, compared to protons, enhanced values of $v_1$ and $v_2$ for light nuclei ($d$, $t$, and $^3$He) were 
observed in prior heavy-ion collision experiments~\cite{prl75_2100,prl74_2464,prc59_884,prl92_072303,npa876_1,prc102_044906,prc94_034908}. 
These measurements suggest that the $v_1$ of heavier nuclei have
more pronounced energy dependences and may carry more direct information on the 
collective motion of nuclear matter.
Recently, the HADES experiment reported the measurements of anisotropic flow of
$p$, $d$ and $t$ from $\sqrt{s_{\rm NN}} = 2.4$ GeV Au+Au collisions~\cite{HADES2020}.
The STAR collaboration observed the atomic mass number ($A$) scaling of light
nucleus $v_2$ for the reduced transverse momentum ($p_{\rm T}$) range of $p_{\rm
T}/A < 1.5$ GeV/$c$ at $\sqrt{s_{\rm NN}} = 7.7 - 200 $
GeV~\cite{prc94_034908}. 
Similar to the number of constituent quark (NCQ) scaling of hadron collective flow~\cite{Molnar_2003}, 
under the assumptions of small $v_n$ and light nucleus formation by nucleon coalescence in momentum space, 
light nucleus collective flow is expected to follow an approximate 
$A$ scaling
\begin{equation}\label{eq_1}
\centering
v_{n}^{A}(p_{\rm T},y)/A \approx v_{n}^{p}(p_{\rm T}/A,y).
\end{equation}
The STAR observation~\cite{prc94_034908} favors nucleon coalescence,
while the true production mechanism of light nuclei in 
heavy-ion collisions is still an open question.
At lower energies, however, the $v_1$ is not negligibly small as reported 
in this paper. Keeping up to $v_1^2$, Eq.~(\ref{eq_1}) for $n=2$ becomes
\begin{equation}\label{eq_2}
v_2^A(p_{\rm T},y)/A \approx v_2^p(p_{\rm T}/A,y) + \frac{A-1}{2}\Big(v_1^{p}(p_{\rm T}/A,y)\Big)^{2}.
\end{equation}

The coalescence model assumes that light nuclei are formed via the
combination of 
nucleons when these nucleons are near each other both in coordinate and momentum
space near the time of kinetic freeze-out~\cite{Butler_1963,Sato_1981,Zhang_2010,Steinherimer_2012}.
Due to the longer passing time of the colliding ions in the few GeV regime, the interference between the 
expanding central fireball and the spectator remnants becomes more significant than at higher energies.
Since flow is strongly affected by the spectators, one expects to gain insight into the collision dynamics and
the nucleon coalescence behavior from the measurements of light nucleus $v_1$ and $v_2$
in the few GeV energy regime.
In this paper, we report the measurements of $v_1$ and $v_2$ as functions of 
particle rapidity ($y$) and transverse momentum ($p_{\rm T}$)
for $d$, $t$, $^3$He, and $^4$He in fixed-target $\sqrt{s_{\rm NN}} =$ 3 GeV
Au+Au collisions at the STAR experiment. 
\section{Experiment and data analysis}
The data used here were recorded in the fixed-target program by the STAR
experiment~\cite{star}. The lab energy of the beam is 3.85 GeV per nucleon,
equivalent to the center-of-mass energy of  $\sqrt{s_{\rm NN}} = 3$ GeV. 
A detailed description of the STAR detector can be found in~\cite{star}. 
The main tracking and particle identification (PID) detectors are the Time
Projection Chamber (TPC)~\cite{tpc} and
the Time-of-Flight (TOF) barrel~\cite{tof} located inside a 0.5 T solenoidal magnetic
field.
For the fixed target configuration, the Au target is 
installed inside the vacuum pipe
200 cm to the west of the TPC center. The TPC covers the full azimuth
and a pseudorapidity range
$0.1 < \eta < 2$, and the TOF covers the range $0.1 < \eta < 1.5$ in the laboratory
frame.
In this paper, the beam direction is defined as positive, and the 
particle rapidity is given in the collision center-of-mass frame.

For each event, the reconstructed primary vertex is required to
be within 2 cm of the target position along the beam axis.
The transverse $x$, $y$ position of the vertex is required to be within
2 cm of the target located at (0, 2) cm.
The event centrality is estimated from the charged-particle  
multiplicity measured in the TPC within $-2 < \eta <0$ with the help of a Glauber Monte Carlo model~\cite{Ray_2008}.

Charged-track trajectories are reconstructed from the measured space point information in the TPC.
In order to select the primary tracks, a requirement of less than
3 cm is applied on their distance of closest approach (DCA) from
the event vertex. To avoid effects from track splitting,
each track should have at least 15 TPC space points, 
and have more than 52\% of the total possible TPC points used in the track
fitting. The TPC reconstruction efficiency is around 80\% for all light nuclei species.

\begin{figure}[htbp]
\centering
\includegraphics[width=8cm]{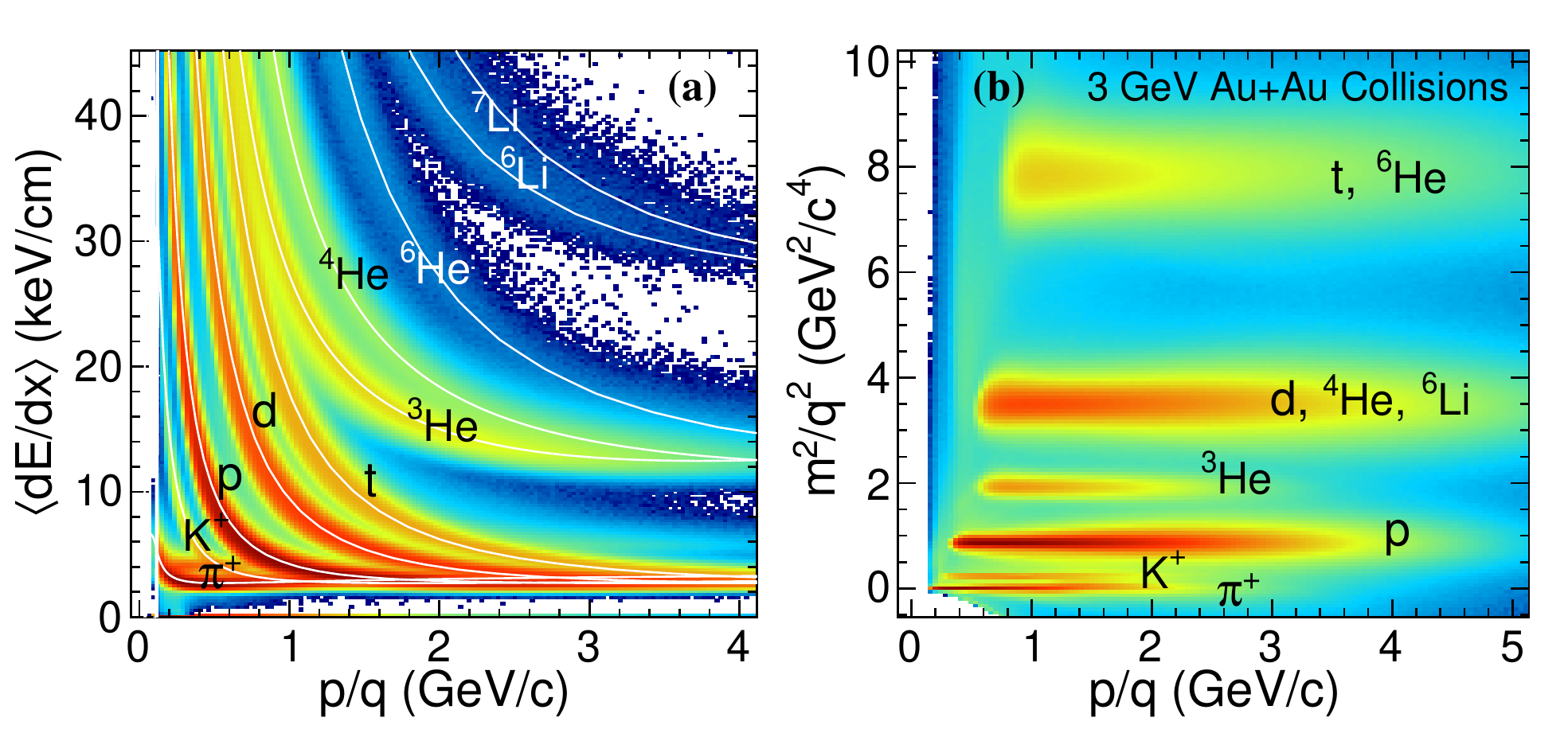}
\caption{(a) The $\langle{dE/dx}\rangle$ of charged tracks versus rigidity in
	Au+Au collisions at $\ems = $ 3 GeV.
The curves are Bichsel expectations for the corresponding particle species as
labeled.
(b) Particle $m^2/q^2$ versus rigidity.
The bands correspond to $\pi^+$, $K^+$, $p$, $^3$He,
$d$, and $t$ as labeled. $^4$He and $^6$Li have the same
$m^2/q^2$ as $d$ and $^6$He has the same $m^2/q^2$ as $t$.
}\label{fig_pid}
\end{figure}

The charged particle identification is accomplished by the specific energy loss $dE/dx$
measured in the TPC. Figure~\ref{fig_pid}a shows the average
$dE/dx$ distribution of charged particles as a function of
rigidity (momentum/charge).
The curves denote the Bichsel expectation for each particle species~\cite{Bichsel}.
At low momenta, the $\langle{dE/dx}\rangle$ bands corresponding to different particle
species are clearly separated
and the particle type can be determined via the variable $z$,
\begin{equation}\label{eq_z}
	z=\ln\biggl(\frac{\langle{dE/dx}\rangle}{\langle{dE/dx}\rangle}_{B}\biggr),
\end{equation}
 where the $\langle{dE/dx}\rangle_{B}$ is the corresponding Bichsel expectation. 
The expected value of $z$ for a given particle type is zero.
At higher momenta, these bands start to overlap. 
A combination of $z$ and $m^2$ of the particle is used to identify the 
high momentum light nuclei with a PID purity higher than 96\%. 
A particle's $m^2$, where $m$ is mass of the particle, is determined by measuring the particle speed using the TOF
system. Figure~\ref{fig_pid}b shows the $m^2/q^2$ distribution as a function of
particle rigidity.

\begin{figure}[htbp]
\centering
\includegraphics[width=7.5cm]{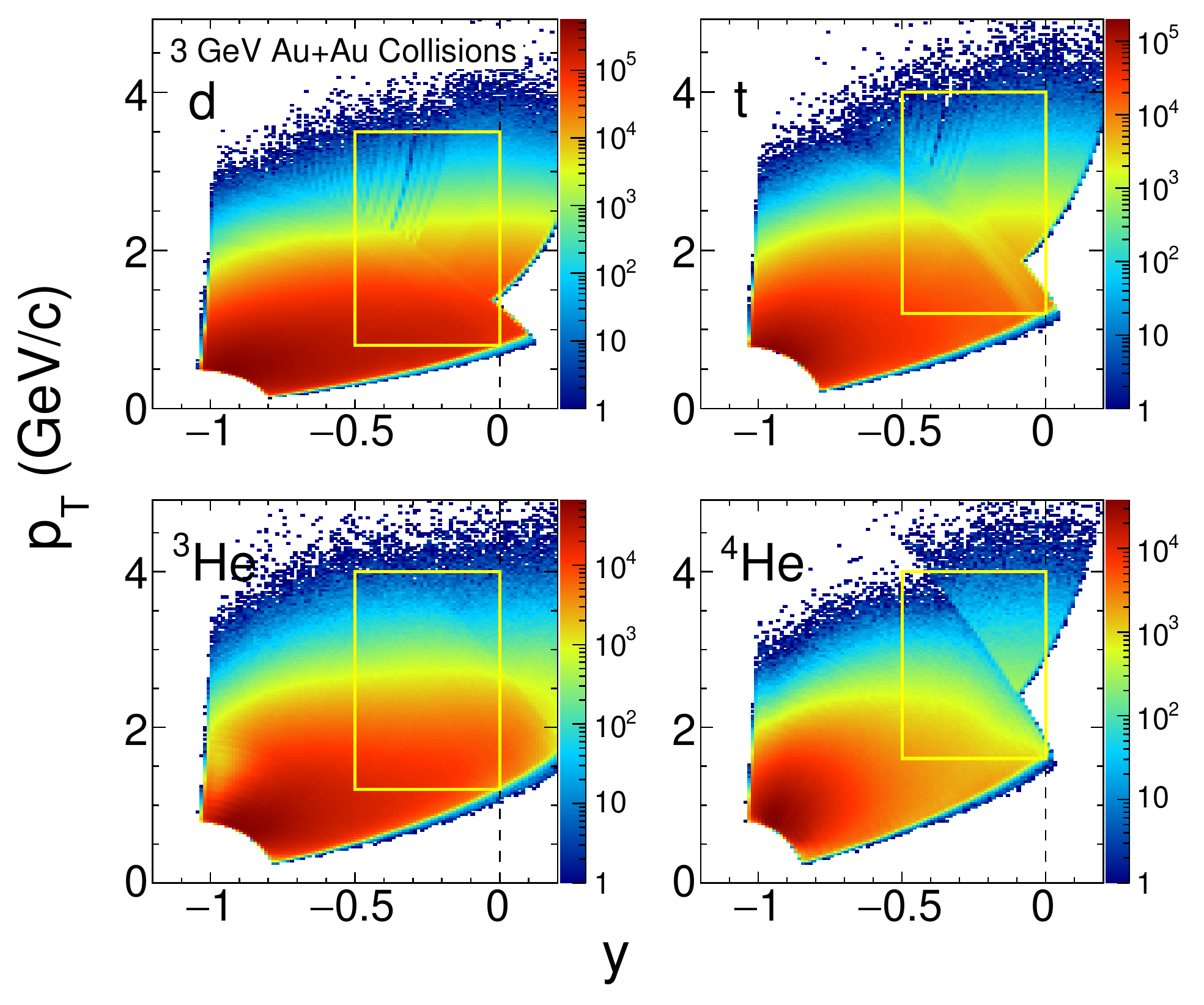}
\caption{The $p_{\rm T}$ versus $y$ acceptances for
	$d$, $t$, $^3$He, and $^4$He at $\ems$ = 3 GeV
	Au+Au collisions. The bands in the distributions are caused by the 
	momentum dependent requirements of the PID. The boxes represent the 
selected phase space for flow calculation.}
\label{fig_pty}
\end{figure}

The proton $v_1$ and $v_2$ are measured over the range of 0.4 $<
p_{\rm T} <$ 2.0 GeV/$c$. In this measurement, the lower cutoffs of light
nucleus
$p_{\rm T}$ are restricted to the same value in terms of $p_{\rm T}/A$ ($>
0.4$ GeV/$c$).
The $p_{\rm T}$ upper limits are determined based on the $p_{\rm T}$ versus $y$
acceptances shown in Fig.~\ref{fig_pty},
within $-0.5 < y < 0$ after each studied light nucleus species is identified. 
The values for $v_1$ and $v_2$ are extracted in the chosen $p_{\rm T}$ ranges:
0.8 $<p_{\rm T} <$ 3.5 GeV/$c$ for $d$, 
1.2 $<p_{\rm T} <$ 4.0 GeV/$c$ for $t$ and $^3$He, and
1.6 $<p_{\rm T} <$ 4.0 GeV/$c$ for $^4$He.
As a result of the limited $\eta$ coverage of the TOF detector, 
within $-0.1 < y < 0$,
the $t$ and $^4$He do not have coverage for $p_{\rm T} <$ 2.1 GeV/$c$
and $p_{\rm T} <$ 2.8 GeV/$c$, respectively.

The coefficients $v_1$ and $v_2$ are determined via a particle's azimuthal angle in momentum
space relative to the azimuth of the reaction plane spanned by the beam
direction and the impact parameter vector. 
While the reaction plane orientation can not be accessed directly in measurements, 
it is common to use the event plane angle to be a proxy of the true
reaction plane~\cite{prc58_1671}.
In this analysis the first-order event plane $\Psi_{1}$ is adopted for both the
$v_1$ and $v_2$ calculations. The $\Psi_{1}$ value
is reconstructed by using information
from the event plane detector (EPD).
A vector 
\begin{equation}
\centering
\vec{Q}=\left(Q_{x}, Q_{y}\right)=\left(\sum_{i} w_{i}\cos(\phi_{i}), \sum_{i}
w_{i}\sin(\phi_{i})\right)
\end{equation} 
is calculated event-by-event. The $\phi_{i}$
is the azimuthal angle of the $i^{\rm th}$ module of the EPD, and its weight 
$w_{i}$ is proportional to the energy deposition. The non-uniformities in the EPD
are corrected by 
subtracting the $\left(\langle Q_{x} \rangle, \langle Q_{y}
	\rangle \right)$
from $\vec{Q}$ in each event~\cite{prc58_1671}, where the angle brackets indicate averaging over all
events. Then the $\Psi_{1}$ is given by
$\Psi_{1}=\tan^{-1}\left(Q_{y}/Q_{x}\right)$. A
shifting method~\cite{prc58_1671} is utilized to make the distribution of the reconstructed  
$\Psi_{1}$ uniform. 

The values $v_{1}$ and $v_{2}$ are computed via $v_{n}=\langle \cos[n(\phi-\Psi_{1})] \rangle/\mathcal{R}_{n}$.
The $p_{\rm T}$- and $y$-dependent reconstruction efficiency of particle tracks is
corrected using a Monte Carlo calculation of simulated particles
embedded into real collision events.
The event plane resolution $\mathcal{R}_{n}$ is determined via a three sub-event plane
correlation method~\cite{prc58_1671}, where the sub-event planes are
reconstructed
separately in different $\eta$ ranges of the EPD and TPC. 
At $\ems=3$ GeV, the resolutions peak in the centrality range 30-40\%
with value of 0.75 and 0.41 for $v_1$ and $v_2$, respectively.

The systematic uncertainties of the measured flow harmonics come from the 
method of selecting charged tracks, from particle identification, and from
the event plane resolution. They are estimated point-by-point
on $v_{1}$ and $v_{2}$ as a function of $y$ and $p_{\rm T}$ for 
each light nucleus species. The systematic
uncertainties arising from the track selection are determined by varying the selection
requirements. The values amount to about 2\% after the statistical fluctuation
effects are removed~\cite{Barlow}. The systematic uncertainties
related to the particle misidentification are determined by varying the
cuts on $z$ and $m^2$, and are found to be 2\% to 8\% depending on the light nucleus
species and their $p_{\rm T}$. 
A common systematic uncertainty arises from the event plane resolution,
and is determined by using combinations of different
$\eta$ sub-events; it is estimated to be less than 2\% and 3\% for $v_1$ and $v_2$, respectively, within the centrality bin 10-40\%.
Additional systematic uncertainty on the $dv_{1}/dy$ slope parameter comes from
the chosen fit range, and is estimated by taking the difference between the fit values
from default range $-0.5 < y < 0$  and from $-0.4 < y < 0$. The typical magnitude
of this systematic uncertainty is found to be 3\% for all light nucleus species. 
In the following figures, the total systematic uncertainty of each data point is represented by the open boxes.

\section{Results and Discussions}
\begin{figure*}[htbp]
\centering
\includegraphics[width=14cm]{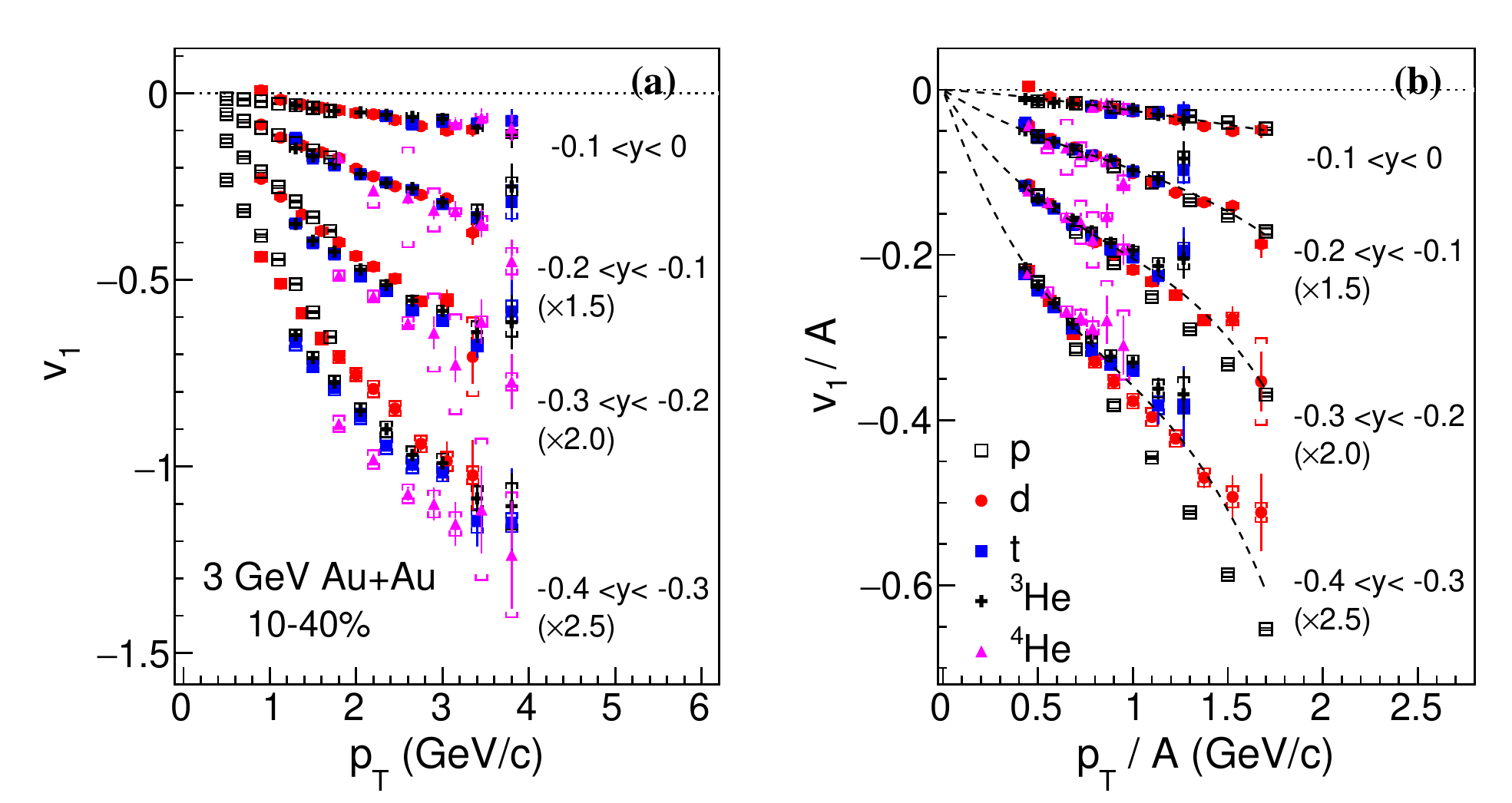}
\caption{(a) The $p_{\rm T}$ and rapidity dependencies of $v_1$ for $p$, $d$, $t$, $^{3}$He, and 
	$^{4}$He in 10-40\% mid-central Au+Au collisions{}
	at$\ems = 3$ GeV. (b) The same results as (a) but both $v_1$ and $p_{\rm T}$
	are scaled by $A$.
	For $t$ and $^4$He, there are no data points at $p_{\rm T}/A < 0.7$
	GeV/$c$ in $-0.1 <y <0$ due to limited acceptance.
	The data points in each rapidity are scaled for clarity. 
	Statistical and systematic uncertainties are represented by vertical lines
	and open boxes, respectively.
	The dashed lines represent the fit to a third-order polynomial function
	of the data
	points to guide the eye. 
}\label{fig_v1pt}
\end{figure*}

The $p_{\rm T}$ dependencies of the light nucleus $v_1$ in different rapidity intervals
are shown in Fig.~\ref{fig_v1pt}. Figure~\ref{fig_v1pt}b shows that the values
of $v_1/A$ of all light nuclei, including protons, approximately follow $A$
scaling for $-0.3 < y < 0$ especially near mid-rapidity. The $v_1$ scaling behavior suggests the light
nuclei are formed via nucleon coalescence in Au+Au collisions at $\ems = 3$ GeV.
The scaling worsens for $p_{\rm T}/A > 1$ GeV/$c$ in the range $-0.4 < y
<-0.3$, where the $v_1$ values are large and the simple coalescence of Eq.~(\ref{eq_1}) may not apply. 
Increasing contamination of target-rapidity ($y=-1.045$) fragments may also play a role.

\begin{figure*}[htbp]
\centering
\includegraphics[width=16cm]{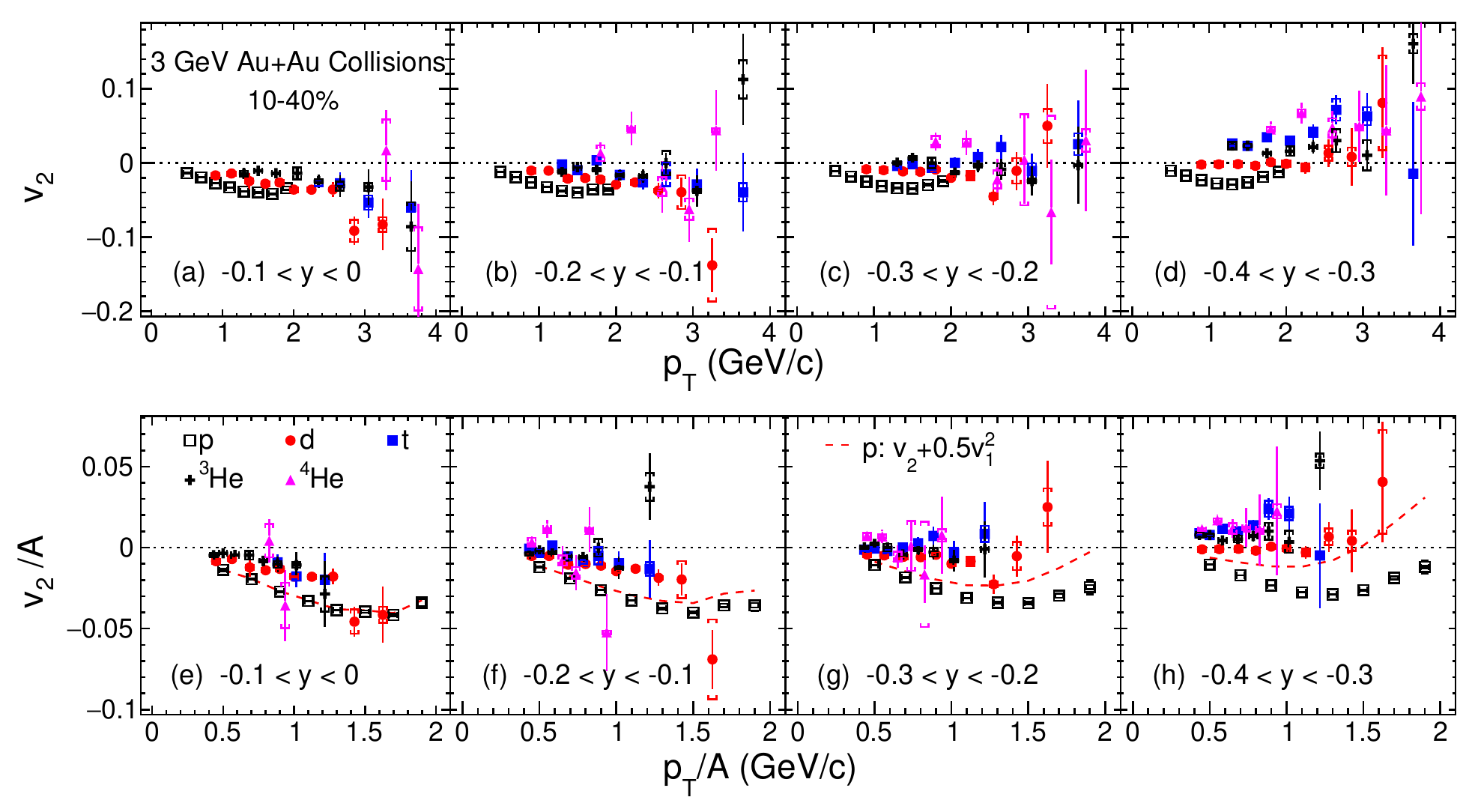}
\caption{
	Upper panels: The $p_{\rm T}$ and $y$ dependencies of $v_{2}$ for $p$, $d$, $t$, $^{3}$He, and 
	$^{4}$He in 10-40\% mid-central Au+Au collisions
	at $\ems = 3$ GeV. Lower panels: The same results as in upper panels but both
	$v_2$ and $p_{\rm T}$ are scaled by $A$. The dashed lines are the $v_2$ expectation 
	for $d$ by Eq.~(\ref{eq_2}).
	Statistical and systematic
	uncertainties are represented by vertical lines and open boxes, respectively.
}\label{fig_v2pt}
\end{figure*}

The upper panels of Fig.~\ref{fig_v2pt} show the dependencies of $v_2$ in different rapidity intervals. 
At mid-rapidity, $-0.1 < y < 0$, the $v_2$ values are negative
for all measured light nucleus species. Moving away from mid-rapidity,
the $v_2$ magnitudes decrease gradually, and become positive for $t$, 
$^{3}$He, and $^{4}$He at larger $p_{\rm T}$, while the $v_2$ of protons and
$d$ remain negative within $-0.4<y<0$. Moreover, the proton $v_2$ has a stronger 
non-monotonic $p_{\rm T}$ dependence compared to other light nuclei. The lower
panels of Fig.~\ref{fig_v2pt} show
$v_2/A$ as a function of $p_{\rm T}/A$ and they do not follow the same trend.
Taking into account the effect of $v_1$ by Eq.~(\ref{eq_2}), the naive momentum 
coalescence expectation of $v_2$ for $d$ is shown in the dashed curves. 
While the $v_1$ effect may partially explain the trend with increasing rapidity,
the $v_2$ data significantly deviate from the curve (shown only for $d$, but
similar behavior is also found for $t$, $^3$He, and $^4$He).
This indicates that no $A$ scaling is observed in these data for light nucleus $v_2$ at $\ems = 3$ GeV. 
The $A$ scaling has been observed for $p_{\rm T}/A <$ 1.5 GeV/$c$ in
higher energy Au+Au collisions at $\ems =7.7 -200$ GeV~\cite{prc94_034908}.
There, as a supporting evidence for the formation of the QGP, 
the $v_2$ of hadrons follow an approximate NCQ scaling~\cite{prl92_052302, plb597_328, prl95_112301}. 

\begin{figure}[htb]
\centering
\includegraphics[width=8cm]{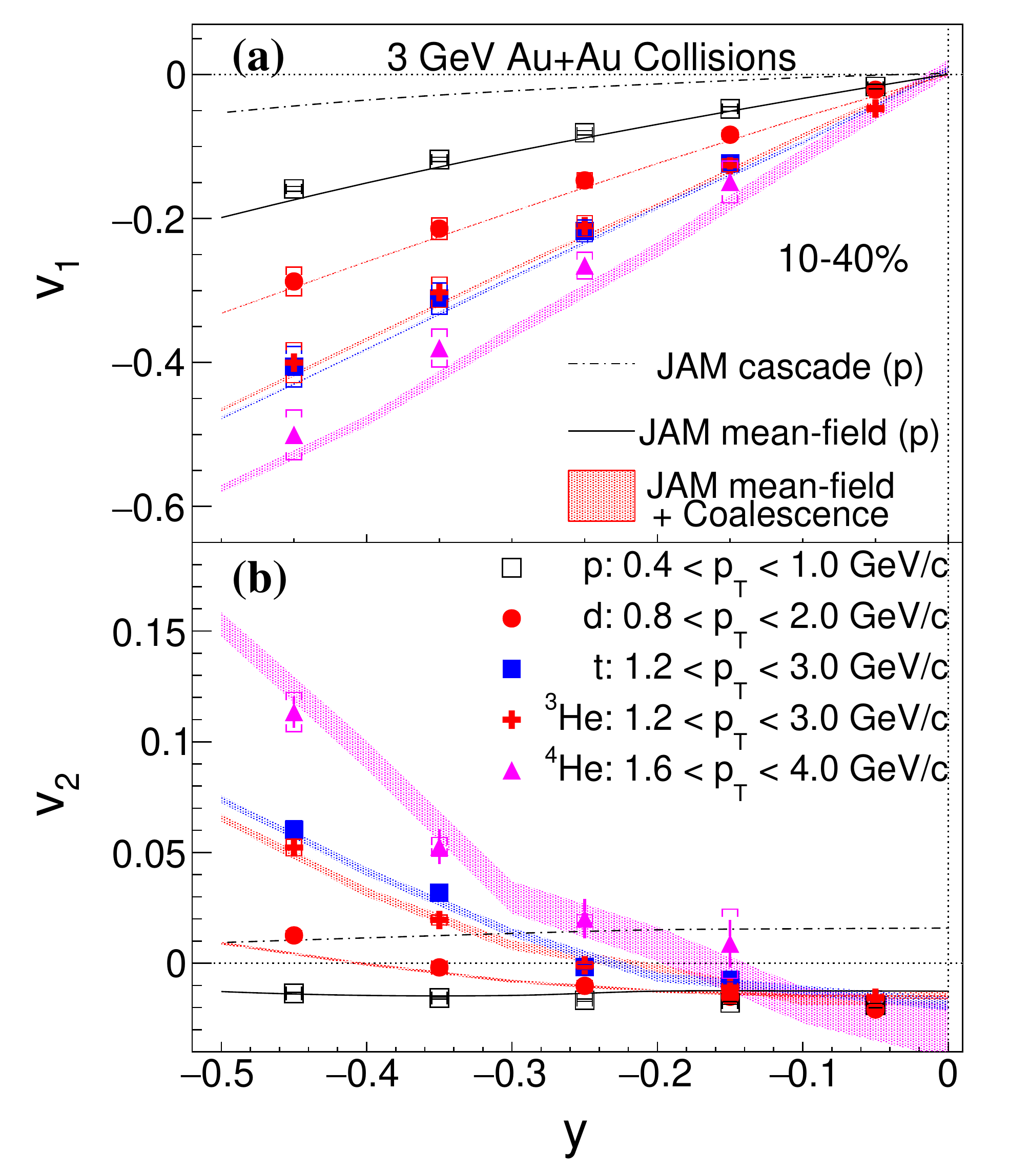}
\caption{
	Rapidity dependencies of light nucleus $v_{1}$ (a) and $v_{2}$ (b)
	in 10-40\% mid-central Au+Au collisions at	$\ems = 3$ GeV. For $t$ and $^{4}$He, the points in $-0.1 < y < 0$ are
	absent due to limited acceptance.
	The dash-dotted line and solid line represent the results for protons from the cascade
	and mean-field modes of JAM, respectively. The bands are
	the results for light nuclei from JAM mean-field plus coalescence
	calculations.
	Systematic uncertainties are represented by open boxes.
}\label{fig_v1v2y}
\end{figure}

Figure~\ref{fig_v1v2y} shows light nucleus $v_1$ and $v_2$ as a function of
rapidity integrated in the chosen $p_{\rm T}$ ranges. There is a clear mass ordering both for $v_1$ and for $v_2$,
namely, the heavier the mass of a nucleus, the stronger the rapidity dependence in
$v_1$ and $v_2$.
At mid-rapidity, $-0.1 <y < 0$, the value of $v_2$ is negative and nearly
identical for $p$, $d$, and $^3$He. 
The negative $v_2$ at mid-rapidity may be caused by shadowing of 
the spectators as their passage time is comparable with the expansion time of
the
compressed system
at $\ems = 3$ GeV~\cite{prl92_072303,npa876_1}. 
During the expansion of the participant zone, the particle emission directed
toward the reaction plane is blocked by the spectators that are still passing the participant zone.  
Moving away from mid-rapidity, the proton $v_2$ remains negative
and those of other light nuclei gradually become positive. 
A similar strong rapidity dependence of light nucleus $v_2$ has also been reported by the HADES experiment~\cite{HADES2020}. 
Nuclear fragmentation may play a role in the production of those light nuclei, the effect of which is 
beyond the scope of the present investigation.

To further understand light nucleus formation and the scaling behavior
of $v_1$ and $v_2$,
we employ a transport model, Jet AA Microscopic Transportation Model
(JAM)~\cite{jam}, to simulate the proton and neutron production 
from the initial collision stage to the final hadron transport in $\ems =$ 3 GeV
Au+Au collisions.
Both the cascade mode and the mean-field mode of JAM calculations are
performed. 
In the cascade mode, particles are propagated as in vacuum (free streaming) between 
collisions with other particles. 
In the mean-field mode~\cite{prc72_064908}, a momentum-dependent potential with 
the incompressibility parameter 
$\kappa=380$ MeV is acting on the nucleon evolution.
The resulting proton $v_1$ and $v_2$  
from the mean-field mode are consistent with the experimental observations (see
solid-lines in Fig.~\ref{fig_v1v2y}).
However, the simulation results from JAM cascade mode underestimate the magnitudes 
of proton $v_1$ and give positive values for proton $v_2$ within $-0.5 < y
<0$, opposite to the data.
Note that the calculations from the mean-field mode, 
which reproduce the observed collectivity of proton and $\Lambda$~\cite{pv1_3gev}, 
impose stronger repulsive interactions among baryons.

The current JAM model does not create light nuclei.
An afterburner, a coalescence approach, is employed to form the light nuclei
using the proton and neutron phase-space distributions at a fixed time of 50 fm/$c$.  
For each nucleon pair, the momentum and position of each nucleon is boosted to the rest
frame of the pair. The relative momentum $\Delta p$ and the relative coordinate 
$\Delta r$ of the two nucleons are evaluated in the rest frame.
If the $\Delta p < $ 0.3 GeV/$c$ and $\Delta r <$ 4 fm, then the nucleon
pair is marked as a $d$~\cite{prc99_014901}. A similar process is used for the
formation of $t$ $(nnp)$, $^3$He $(npp)$ and $^4$He $(nnpp)$, where the
constituent nucleons are added one by one according to the $\Delta p$ and
$\Delta r$ in the rest frame of the nucleon and a light nucleus core. The resulting
light nucleus $v_1$ and $v_2$, as functions of rapidity, are shown as
bands in Fig.~\ref{fig_v1v2y}a and \ref{fig_v1v2y}b, respectively.
Qualitatively both dependencies are well reproduced by the mean-field mode of the JAM plus coalescence
calculations. It is noteworthy that the sign change in $v_2$ of protons
(negative) compared to light nuclei (positive) with increasing rapidity is also reproduced by the model calculations.
Note, the broken $A$ scaling for light nucleus $v_2$ is consistent with the nucleon coalescence picture. 
On the other hand, the cascade mode of the JAM cannot reproduce the measured $v_1$ and $v_2$
of protons, as shown by the dash-dotted curves in Fig.~\ref{fig_v1v2y}. 
As a result, calculations with JAM cascade plus coalescence fail to
reproduce the $y$ dependence of $v_1$ and $v_2$ of light nuclei.

\begin{figure}[htb]
\centering
\includegraphics[width=8.0cm]{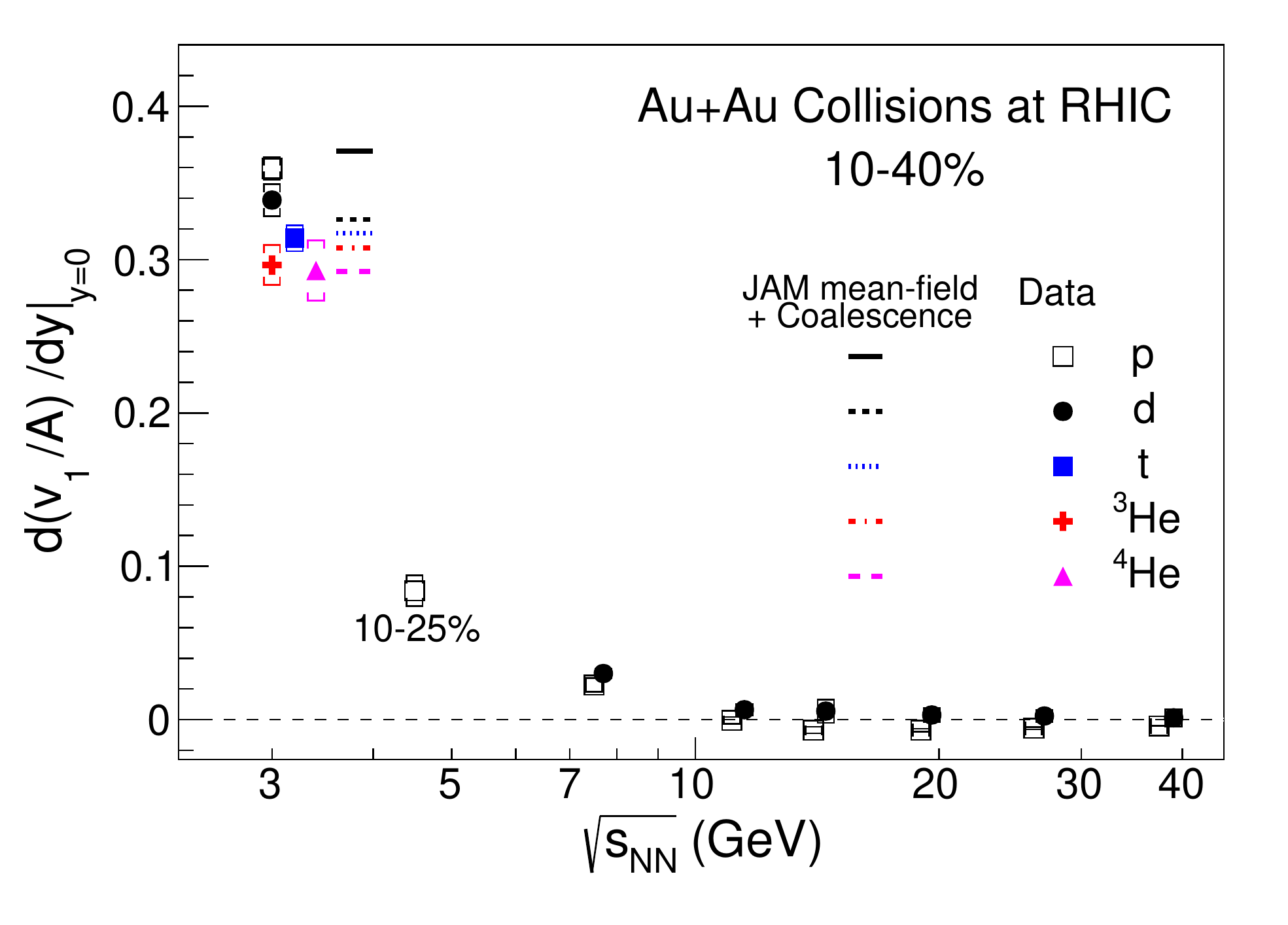}
\caption{
	Light nucleus scaled $v_{1}$ slopes $\big(d(v_1/A)/dy|_{y=0}\big)$ as a
	function of collision energy in 10-40\% mid-central Au+Au collisions. Statistical and systematic
	uncertainties are represented by vertical lines and open boxes, respectively. The data
	points above 7 GeV are taken from~\cite{prc102_044906}. The proton
	result at $\ems = 4.5$ GeV is for 10-25\% Au+Au collisions~\cite{prc103_034908}.	
	For clarity, the data points are shifted horizontally. Results of the JAM
	model in the mean-field mode plus coalescence calculations are shown as color bars.  
}\label{fig_dv1dy}
\end{figure}

A first order polynomial function
is employed to fit $v_1$ in Fig.~\ref{fig_v1v2y}a within rapidity range $-0.5 < y <0$.
The extracted slope parameters, $dv_1/dy$, scaled by $A$, for light nuclei are shown in
Fig.~\ref{fig_dv1dy} as functions of the collision energy, together
with existing data from higher energies.
The values of $(dv_1/dy)/A$ at 3 GeV for all measured light nuclei are positive and grouped
together with that of the protons. The results of the JAM model in mean-field mode plus coalescence
calculations for $p$, $d$, $t$, $^3$He and $^4$He in 3 GeV Au+Au collisions are also shown with
corresponding bars. The same experimental cuts have been applied in the
calculations and the resulting slope parameters are consistent with the data
including the relative order. 
The agreement between experimental data and model calculations implies that at 3 GeV 
these light nuclei are formed via the coalescence processes and baryonic interactions dictate
their dynamics.     

At higher collision energies, the $v_1$ of $d$ has been
measured from $\ems = 7.7 - 39$ GeV Au+Au collisions by the STAR
experiment~\cite{prc102_044906}. At $\ems =  7.7$ GeV, the $v_1$ slope of $d$ 
follows $A$ scaling within the statistical and systematic uncertainties.
For energy $\ems > 7.7$ GeV, the value of proton $dv_1/dy$ is negative and the
corresponding $v_1$ slopes of $d$ are positive with larger uncertainties. The
different scaling behavior of light nuclei $dv_1/dy$ at $\ems \le 7.7$ GeV and
$\ems >11.5$ GeV may indicate a different production mechanism. At higher
energies where a QGP is formed, the dominant interactions are partonic in nature.
At 3 GeV, baryonic interactions are likely dominant and light nuclei may primarily be 
formed via coalescence of nucleons. 
Fragmentation contribution may also play a role which requires further investigation.

\section{Summary}
In summary, we present the directed flow $v_1$ and elliptic flow $v_2$ of $d$, $t$, $^3$He,
and $^4$He for 10-40\% centrality in Au+Au collisions at $\ems = $ 3 GeV. 
The light nucleus $v_1$, as function of both transverse momentum and particle
rapidity, follow an approximate atomic mass number $A$ scaling at rapidity $-0.5 < y < 0$,
consistent with the nucleon coalescence model calculations.
On the other hand, the light nucleus $v_2$ do not follow the simple $A$ scaling, 
even after taking into account the contribution from the comparable magnitude of $v_1^2$.
At mid-rapidity $-0.1 <y <0$, the value of $v_2$ is negative for all light nuclei, implying a
shadowing effect due to the longer passage time of the spectators. Away
from the mid-rapidity, the values of light nucleus $v_2$
become positive and the corresponding proton $v_2$ remains negative.
The JAM model, with the baryon mean-field (incompressibility parameter $\kappa$ = 380 MeV and a momentum dependent potential), 
and a nucleon coalescence qualitatively reproduce both the $v_1$ and $v_2$ as
functions of rapidity for all reported 
light nuclei. 
On the other hand, the results from the JAM cascade mode plus coalescence fail to describe the
data.
Our results suggest that the light nuclei are likely formed via the
coalescence of nucleons at $\ems = $ 3 GeV Au+Au collisions, where baryonic interactions
dominate the collision dynamics.


\section*{Acknowledgement}
We thank the RHIC Operations Group and RCF at BNL, the NERSC Center at LBNL, and the Open Science Grid consortium for providing resources and support.  This work was supported in part by the Office of Nuclear Physics within the U.S. DOE Office of Science, the U.S. National Science Foundation, the Ministry of Education and Science of the Russian Federation, National Natural Science Foundation of China, Chinese Academy of Science, the Ministry of Science and Technology of China and the Chinese Ministry of Education, the Higher Education Sprout Project by Ministry of Education at NCKU, the National Research Foundation of Korea, Czech Science Foundation and Ministry of Education, Youth and Sports of the Czech Republic, Hungarian National Research, Development and Innovation Office, New National Excellency Programme of the Hungarian Ministry of Human Capacities, Department of Atomic Energy and Department of Science and Technology of the Government of India, the National Science Centre of Poland, the Ministry  of Science, Education and Sports of the Republic of Croatia, RosAtom of Russia and German Bundesministerium f\"ur Bildung, Wissenschaft, Forschung and Technologie (BMBF), Helmholtz Association, Ministry of Education, Culture, Sports, Science, and Technology (MEXT) and Japan Society for the Promotion of Science (JSPS).



\end{document}